\documentclass[aps,prl,twocolumn,amsmath,amssymb]{revtex4}
\usepackage{graphicx}
\usepackage{Jonasmacros}
\usepackage{bm}
\begin{document}
\date{\today}
\title{Spin Inelastic Electron Tunneling Spectroscopy on Local Spin Adsorbed on Surface}
\author{J. Fransson}
\email{Jonas.Fransson@fysik.uu.se}
\affiliation{Department of Physics and Materials Science, Uppsala University, Box 530, SE-751 21 Uppsala, Sweden}

\begin{abstract}
The recent experimental conductance measurements taken on magnetic impurities on metallic surfaces, using scanning tunneling microscopy technique and suggesting occurrence of inelastic scattering processes, are theoretically addressed. We argue that the observed conductance signatures are caused by transitions between the spin states which have opened due to e.g. exchange coupling between the local spins and the tunneling electrons, and are directly interpretable in terms of inelastic transitions energies. Feasible measurements using spin-polarized scanning tunneling microscopy that would enable new information about the excitation spectrum of the local spins are discussed.
\end{abstract}
\maketitle

Inelastic scattering processes play a crucial role in studies of excitation spectrums of e.g. single atoms and clusters of atoms, molecules, quantum dots, carbon nanotubes, and graphene. Signatures of inelastic scattering are often displayed in conductance measurements as steps that cannot be ascribed to the resonant elastic tunneling, something which led to the discovery of molecular vibrations in metal-insulator-metal junctions exposed to propionic acid \cite{jaklevic1966}. Inelastic features arising from vibrations have since been studied in e.g. molecular structures \cite{park2000}, Josephson junctions \cite{gorelik2001}, and surfaces \cite{gawronski2008}.

Recently, inelastic scattering have been observed in many studies of magnetic structures. Magnetic anisotropy of magnetic atoms and clusters have, for instance, been studied through spin inelastic scattering \cite{gambardella2003,heinrich2004,hirjibehedin2006,hirjibehedin2007,balashov2008}. Signatures of spin inelastic scattering processes are, furthermore, frequently appearing in excitation spectrums taken on bulk materials, e.g. Fe, and thin films, e.g. Co \cite{balashov2006}, and individual magnetic adatoms, e.g. Fe and Cr \cite{wahl2007}, Co \cite{yayon2007,meier2008}, and molecular magnets \cite{chen2008}. In contrast, there are only a few theoretical studies in which spin inelastic scattering processes are included in a non-equilibrium description \cite{fransson2008,fransson2009,rossier2009}.

Here, we address the theoretical description of inelastic signatures in the (differential) conductance through a tunneling barrier in which a local spin is embedded, and we include the possibility of magnetic source and drain. The set-up is, thus, pertinent for scanning tunneling microscope with or without spin-polarized tip. As we discuss below, the conductance can be naturally divided into three terms of which the last is sensitive to the spin fluctuations in the local spin structure. Those fluctuations include transitions between the ground state and excited states which are forbidden in the absence of interactions between the local spins and the surrounding electrons in e.g. tunneling current and substrate. The coupling between these states, which is mediated by the de-localized electrons, opens new channels for the tunneling conductance which subsequently appear as steps in the resulting conductance measurement.

In this Letter, we describe the STM setup in terms of electrons tunneling between a tip and a substrate onto which a sample is located. In our case the sample comprise a spin moment which may constitute one or several spins. We write $\bfS=\sum_n\bfS_n$, where the spin $\bfS_n$ is located at $\bfr_n$. The tunneling between the tip, $\Hamil_\tip=\sum_{\bfp\sigma}\leade{\bfp}\cdagger{\bfp}\cc{\bfp}$, and the substrate, $\Hamil_\sub=\sum_{\bfk\sigma\sigma'}\dote{\bfk\sigma\sigma'}\cdagger{\bfk}\cs{\bfk\sigma'}$, occurs in presence of the local spin moments, and we use the model $\Hamil_T=\sum_{\bfp\bfk\sigma\sigma'n}\cdagger{\bfp}\hat{T}_{\sigma\sigma'}(\bfr_n,t)\times \cs{\bfk\sigma'}e^{i\bfk\cdot\bfr}+H.c.$. Here $\hat{T}(\bfr_n,t)=T_0+T_n\bfsigma\cdot\bfS_n$ \cite{balatsky2002}, where $T_0$ and $T_n$ give the tunneling rates for electrons not interacting and interacting, respectively, with the local spins. We include the possibility of magnetic tip and substrate. However, since we do not require them to be collinear with one another, we here have taken the spin quantization axis of the tip as the global reference frame. Due to the local nature of the spins, we have included a spatial dependence in the interacting tunneling rate, and we use e.g. $T_n=T'\exp{(-|\bfr-\bfr_n|/\lambda)}$, where $\lambda$ is the decay length. Typically, from expansion of the work function for tunneling $T'/T_0\sim J/U$ \cite{balatsky2002}, where $J$ and $U$ are the spin-spin exchange interaction parameter and spin-independent tunneling barrier, respectively. For metals and semiconductors we may use $T'/T_0\sim0.1$ \cite{bhattacharjee1992}. Here, also $\bfsigma=(\sigma^x,\sigma^y,\sigma^z)$ are the Pauli matrices.

We note that including a Kondo-like (exchange) interaction, e.g. $-J_K\sum_{\bfk\bfk'\sigma\sigma'}\cdagger{\bfk}\bfsigma_{\sigma\sigma'}\cdot\bfS_n\cs{\bfk'\sigma'}$, between the local spins and the surface electrons in the substrate does not qualitatively alter picture for the spin moment. This can be understood since this interaction affects the local spin in a physically similar way as the interaction between the local spin and the tunneling current. We therefore omit this interaction in favor of establishing a comprehensive picture of the inelastic signatures measured in the STM measurements.

Using non-equilibrium technique on the Keldysh contour $C$, we derive the tunneling current $I(t)$ from the fundamental expression
\begin{align}
I(t)=&
	-e\frac{d}{dt}\sum_{\bfp\sigma}\av{\cdagger{\bfp}(t)\cc{\bfp}(t)}
\nonumber\\=&
	-\frac{2e}{\hbar}\im\sum_{\bfp\bfk\sigma\sigma'}
		\av{\cdagger{\bfp}(t)\hat{T}_{\sigma\sigma'}(\bfr_n,t)\cs{\bfk\sigma'}(t)}e^{i\bfk\cdot\bfr}.
\end{align}
Expanding the average in the last expression according to $\av{A(t)}\approx(-i)\int_C\av{\com{A(t)}{\Hamil_T(t')}}dt'$, and by converting to real times, we find that the current can be written
\begin{widetext}
\begin{align}
I(t)=
	\frac{2e}{\hbar}\re\sum_{\bfp\bfp'\bfk\bfk'}\sum_{\sigma\sigma'\sigma''\sigma'''}
		\int_{-\infty}^t
		\langle[\cdagger{\bfp}(t)\hat{T}_{\sigma\sigma'}(\bfr_n,t)\cs{\bfk\sigma'}(t),
			\csdagger{\bfk'\sigma'''}(t')\hat{T}_{\sigma'''\sigma''}(\bfr_m,t')\cs{\bfp'\sigma''}(t')]
		\rangle
			e^{i(\bfk-\bfk')\cdot\bfr}dt'.
\end{align}
\end{widetext}
By expanding the tunneling operator $\hat{T}$ into its components $T_0$ and $T_1$, we find that the current can be naturally written as a sum of three terms, i.e. $I(t)=\sum_{i=0}^2I_i(t)$. We consider the differential conductance $dI/dV=\sum_{i=0}^2dI_i/dV$ of stationary source drain voltages since such conditions are predominant in experimental situations. The first two contributions to the conductance were discussed in detail in Ref. \cite{fransson2009} and will, therefore, not be further considered here. In the stationary regime and at low temperatures, we can write the last contribution to the conductance according to
\begin{widetext}
\begin{align}
\frac{dI_2(\bfr,V)}{dV}=&
	i\biggl(\frac{\pi}{2}\biggr)^2\frac{\sigma_0}{2}
	\sum_{nm}T_nT_m
	\int
		\Big(
			f(\dote{})\delta(\dote{}+eV-\omega)+[1-f(\dote{})]\delta(\dote{}+eV+\omega)
		\Bigr)
\nonumber\\&\times
	\Bigl(
		8[n(\dote{})N(\bfr,\dote{F})+\bfm(\dote{})\cdot\bfM(\bfr,\dote{F})]
		\chi^z_{nm}(\omega)
\nonumber\\&
		+[n(\dote{})N(\bfr,\dote{F})-\bfm(\dote{})\cdot\bfM(\bfr,\dote{F})]
		[\chi^{-+}_{nm}(\omega)+\chi^{+-}_{nm}(\omega)]
\nonumber\\&
		+[m_z(\dote{})N(\bfr,\dote{F})
		-n(\dote{})M_z(\bfr,\dote{F})\cos\theta]
		[\chi^{-+}_{nm}(\omega)-\chi^{+-}_{nm}(\omega)]
	\Bigr)
	\frac{d\omega}{2\pi}d\dote{}.
\label{eq-I2}
\end{align}
\end{widetext}
Here, $\sigma_0=2e^2/h$ is the fundamental conductance unit and $f(\dote{})$ is the Fermi function, whereas $n(\dote{})$  ($\bfm(\dote{})$) and $N(\bfr,\dote{})$ ($\bfM(\bfr,\dote{})$) are the electronic (magnetic) densities in the tip and substrate, respectively, with the angle $\theta$ between $\bfm$ and $\bfM$, i.e. $\bfm\cdot\bfM=|\bfm||\bfM|\cos\theta$. We have defined the spin-spin correlation functions e.g. $\chi^{-+}_{nm}(\omega)=\int\chi^{-+}(t)e^{-i\omega t}dt$ and $\chi^z_{nm}(\omega)=\int\chi^z_{nm}(t)e^{-i\omega t}dt$ \cite{SSassumption}, where $\chi^{-+}_{nm}(t,t')=(-i)\av{S_n^-(t)S_m^+(t')}$ and $\chi^z_{nm}(t,t')=(-i)\av{S^z_n(t)S^z_m(t')}$, and the bias voltage $V$ across the junction.



The conductance given in Eq. (\ref{eq-I2}), provides a contribution to the total conductance in which signatures from the spin-spin correlations, or spin fluctuations, are present. First it is important to notice that $dI_2/dV$ is finite for any polarization of the tip and substrate, even when both are non-magnetic. Hence, under any condition, this conductance depends on fluctuations of the local spin moments, here expressed by the correlation functions $\chi_{nm}^{-+}(\omega)$, $\chi_{nm}^{+-}(\omega)$, and $\chi_{nm}^z(\omega)$.

The main contribution to the differential conductance $dI/dV$ is generated by the elastic tunneling processes, which are captured by the first contribution $dI_0/dV$. The steps seen in the STM conductance experiments of local spins, c.f. e.g. \cite{hirjibehedin2006,hirjibehedin2007,wahl2007,yayon2007,meier2008,chen2008}, cannot be related to elastic processes, because of the absence of local electron levels that would resonate with the bias voltage at those energies. Instead, the steps appearing in the $dI/dV$ can be ascribed to inelastic scattering processes which open new channels for conductance. In the present case, such steps are attributed to transitions between different spin states.

We may think of e.g. a local spin moment $\bfS$ comprising two coupled spins $\bfS_n$, $n=1,2$, a spin dimer, and consider them to be anti-ferromagnetically coupled. The ground state is a spin singlet $\ket{S=0,\sigma=0}$, while the first excited states constitute a spin triplet $\ket{S=S_1+S_2,\sigma=0,\pm1}$. Assuming an exchange energy $|J|=|E_S-E_T|>k_BT$, where $E_{T(S)}$ denotes the triplet (singlet) energy, in order to prevent thermal excitations at zero bias, the equilibrium conductance is given by the elastic tunneling between the tip and the substrate only, i.e. $dI/dV=dI_0/dV$. Effects from tunneling electrons scattering off the local spin moment averages to zero. 

The coupling to the tunneling electrons via the spin-spin interaction e.g. $\cdagger{\bfp}\bfsigma_{\sigma\sigma'}\cdot\bfS_n\cs{\bfk\sigma}$ enables, on the other hand, each individual spin constituting $\bfS$ to undergo spin-flip transitions which are assisted by spin-flips of the tunneling electrons. Due to this coupling, the correlation function e.g. $\bfsigma_{\sigma\sigma'}\cdot\av{\bfS_n(t)\bfS_m(t')}\cdot\bfsigma_{\sigma''\sigma}$ is non-vanishing, in general. The spin-spin interaction, thus, provides a coupling between the singlet and triplet states which supports transitions between them. As a result of these transitions, a new channel for conductance opens at bias voltages $V\geq|J|/e$.

We write the Hamiltonian for the spin $\bfS_n$ according to $\Hamil_n=g\mu_B\bfB\cdot\bfS_n$, where $\bfB$ is an external magnetic field. We account for the effective exchange interaction between the spin moments by a Heisenberg model $\Hamil_J=-J\sum_{n\neq m}\bfS_n\cdot\bfS_m$. This effective exchange comprises a combination of e.g. direct Heisenberg exchange and RKKY-like (Ruderman-Kittel-Kasuya-Yosida) exchange. The sign of the effective exchange parameter, $J$, may vary with distance between the spins in the cluster \cite{meier2008}. In this way we can describe clusters of spins with a total spin moment $\bfS$. We define the resulting eigensystem $\{E_\sigma,\ket{S,\sigma}\}_{\sigma=-S}^S$, for the eigenenergies and eigenstates, respectively, of the model $\Hamil_S=\sum_n\Hamil_n+\Hamil_J+\sum_n[D(S_n^z)^2+E\{(S_n^x)^2-(S_n^y)^2\}]$, where we have added the spin anisotropy fields $D$ and $E$.

The spin-spin correlation functions $\chi^{-+}_{nm}(\omega)$, $\chi^{+-}_{mn}(\omega)$, and $\chi^z_{nm}(\omega)$ are calculated in terms of the eigensystem for the total spin, giving
\begin{subequations}
\label{eq-chi}
\begin{align}
	\chi_{nm}^{\mp\pm}(\omega)=&
	(-i)2\pi\sum_{iv}
		\bra{i}S_n^\mp\ket{v}\bra{v}S_m^\pm\ket{i}
\nonumber\\&\times
		P(E_i)[1-P(E_v)]\delta(\omega+E_i-E_v),
\label{eq-chimp}\\
	\chi_{nm}^z(\omega)=&
	(-i)2\pi\sum_{iv}
		\bra{i}S_n^z\ket{v}\bra{v}S_m^z\ket{i}
\nonumber\\&\times
		P(E_i)[1-P(E_v)]\delta(\omega+E_i-E_v),
\label{eq-chiz}
\end{align}
\end{subequations}
where $i\ (v)$ denotes the initial (intermediate) state, whereas $P(x)$ accounts for the population in the corresponding state. Eq. (\ref{eq-chi}) provides the general qualitative features of the spin-spin correlation function, and shows that there will appear steps in $dI/dV$ whenever the bias voltage matches the transition energy $\pm(E_i-E_v)$. In the case of a spin dimer, for instance, there appear steps in $dI/dV$ when the bias voltage supports the inelastic transitions between the single and triplet states. Indeed, in the case of non-magnetic tip and substrate the present theory reproduces the results presented in \cite{rossier2009}, and shows an excellent agreement with the experiments in \cite{hirjibehedin2006,hirjibehedin2007}.

The forms used here to describe the spin fluctuations given in Eq. (\ref{eq-chi}) correspond to the atomic limit result of the spin-spin correlation functions, i.e. influence from the tunneling current on the spin fluctuations are omitted. While this expression is a simplification, it contains the central mechanisms which are necessary in order to address the inelastic features that we are considering in this Letter. A more detailed description of the spin-spin correlation functions provides e.g. the widths (lifetimes) of the states involved in the transition processes, whereas the qualitative behavior arising from the spin-spin correlations are captured by the expressions in Eq. (\ref{eq-chi}).

\begin{figure}[tb]
\includegraphics[width=8.5 cm]{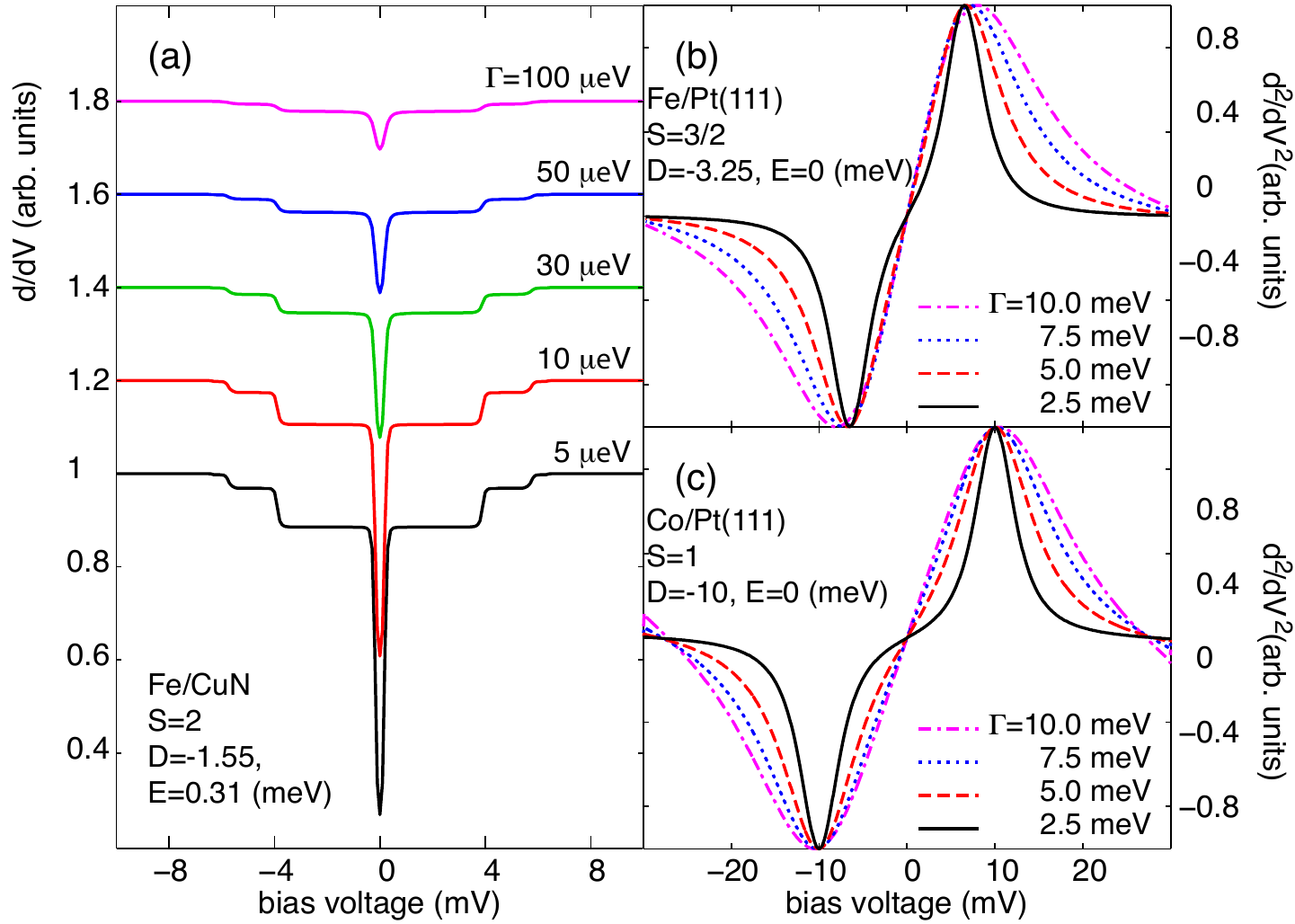}
\caption{(Color online) $dI/dV$ for Fe/CuN (a) at $T=0.5$ K, plots off-set for clarity, $d^2I/dV^2$ for (b) Fe/Pt(111) and (c) Co/Pt(111) for different widths $\Gamma$, at $T=4.3$ K.}
\label{fig-FeCo}
\end{figure}

\begin{table}[b]
\caption{Anisotropy parameters, $D$ and $E$, used for the conductance plots given in  \ref{fig-FeCo}, the best widths, $\Gamma$, for the intermediate states in the spin-spin correlation functions.}
\label{tab-anis}
\begin{tabular}{ccccc}
\hline\hline
& S & D (meV) & E (meV) & $\Gamma$ (meV) \\\hline
Fe/CuN & 2 & -1.55 & 0.31 & $0.01 - 0.03$ \\
Fe/Pt(111) & 3/2 & -3.25 & 0 & $5$ \\
Co/Pt(111) & 1 & -10 & 2 & $10$ \\
\hline\hline
\end{tabular}
\end{table}

Signatures of the excitations in the experimental measurements do have a finite width, which corresponds to that the intermediate states have finite lifetimes. Replacing the delta functions in Eq. (\ref{eq-chi}) by Lorentzian funtions $1/(x^2+\Gamma^2)$, we phenomenologically include the (uniform) lifetime $\hbar/\Gamma$ for all intermediate states. In case of a single Fe on CuN with spin $S=2$ \cite{hirjibehedin2007}, and the anisotropy parameters given in \ref{tab-anis}, we find that the Fe is weakly coupled ($\Gamma\sim10 - 30\ \mu$eV) to the Cu(100) through the CuN layer, see  \ref{fig-FeCo} (a). This value is extracted by comparing the ratio between the maximal and minimal conductance with the experimentally result ($\sim 1/2$). Similarly, we also extract the widths of the single Fe $(S=3/2)$ and Co $(S=1)$ adsorbed onto Pt(111) surface \cite{balashov2008}. In  \ref{fig-FeCo} (b), we have plotted $d^2I/dV^2$ for Fe (upper) and Co (lower) and we find the best correspondence with experiments using the parameters given in \ref{tab-anis}.

In the calculations we have estimated the population numbers $P(E_{i(v)})$ of the states $\ket{i(v)}$ involved in the spin-spin correlation functions, c.f Eq. (\ref{eq-chi}), by making use of the following observation. By expanding the correlation function e.g. $\chi^{+-}_{nm}(t,t')$ in terms of its eigenstates, we can write it as $\chi_{nm}^{+-}(t,t')=\sum_{iv}\av{i|S_n^-|v}\av{v|S^+_m|i}(-i)\av{(\ddagger{i}\dc{v})(t)(\ddagger{v}\dc{i})(t')}$ where $\ddagger{i}$ ($\dc{i}$) creates (annihilates) a particle in the state $\ket{i}$. In the atomic limit, we can employ the decoupling $\av{(\ddagger{i}\dc{v})(t)(\ddagger{v}\dc{i})(t')}=\av{\ddagger{i}(t)\dc{i}(t')}\av{\dc{v}(t)\ddagger{v}(t')}=P(E_i)[1-P(E_v)]e^{i(E_i-E_v)(t-t')}$, where the population number $P(E_i)=\av{\ddagger{i}\dc{i}}$ can be estimated by using e.g. the Gibbs distribution $P(E)=e^{-\beta E}/\sum_ie^{-\beta E_i}$.

The rest of this Letter is devoted to inelastic spectroscopy using SP-STM. It is clear from the previous discussion that the conductances $dI_0/dV$ and $dI_1/dV$ only provide a change in the overall amplitude when the spin-polarization of the tunneling current is varied. We therefore focus the following discussion to the spin-spin correlation function, Eq. (\ref{eq-chi}), in the last contribution of $dI/dV$.

We observe in Eqs. (\ref{eq-I2}) and (\ref{eq-chi}) that the transition sequences $\bra{i}S_n^+\ket{v}\bra{v}S_m^-\ket{i}$, $\bra{i}S_n^-\ket{v}\bra{v}S_m^+\ket{i}$, and $\bra{i}S_n^z\ket{v}\bra{v}S_m^z\ket{i}$ are associated with different projections of the spin-resolved LDOS in the tip and substrate. The sum and difference of the first two sequences couple to $n(\dote{})N(\dote{F})-\bfm(\dote{})\cdot\bfM(\dote{F})$ and $m_z(\dote{})N(\dote{F})-n(\dote{})M_z(\dote{F})\cos\theta$, respectively, while the last sequence couples to $n(\dote{})N(\dote{F})+\bfm(\dote{})\cdot\bfM(\dote{F})$. These different couplings reflect a possibility to enhance or attenuate the response of certain inelastic transitions by using different combinations of electronic and magnetic densities in the tip and substrate.

In STM without spin-polarization ($\bfm,\bfM=0$), for instance, the excitation spectrum can be analyzed to certain detail by means of applying an external magnetic field e.g. along the $z$-direction, or the spin quantization axis of the sample, which introduces a Zeeman split of the levels. In order to be concrete, consider a single adatom with $S=1$, and anisotropy fields $D<0$ and $E=0$, which is described by the states $\ket{S=1,\sigma=0,\pm1}$, where $E_\sigma<E_0$ at $\bfB=0$. In this system, the leading contribution to the term in $dI_2/dV$ containing $\chi_{nm}^{-+}(\omega)+\chi_{nm}^{+-}(\omega)$, c.f Eq. (\ref{eq-I2}), is proportional to $\sum_{\sigma=\pm1}[P(E_\sigma)-P(E_0)][f(eV-E_\sigma+E_0)-f(eV+E_\sigma-E_0)]$, while the term containing $\chi_{nm}^{-+}(\omega)-\chi_{nm}^{+-}(\omega)$ vanishes. The former contribution generates steps in the conductance at bias voltages $eV=\pm(E_\sigma-E_0)$, separated by $|E_{+1}-E_{-1}|$, whenever the states $\ket{1,0}$ and $\ket{1,\sigma}$ are unequally occupied. Hence, a spin split imposed by an external magnetic field is detectable through separate but equally high steps in the conductance.

Using SP-STM opens further possibilities in the studies of spin systems, since then the term in $dI_2/dV$ containing $\chi_{nm}^{-+}(\omega)-\chi_{nm}^{+-}(\omega)$ is finite. In case of the $S=1$ adatom, the leading contribution to this term is proportional to $\sum_{\sigma=\pm1}\sigma[P(E_\sigma)+P(E_0)-2P(E_\sigma)P(E_0)][f(eV-E_\sigma+E_0)-f(eV+E_\sigma-E_0)]$. Notice that the contributions to this term have opposite signs. Hence, by combining the electronic and magnetic densities in the tip and substrate such that $m_zN-n\bfM_z\cos\theta<0$, this term leads to an attenuated (intensified) signal from the transition $\ket{1,0}\bra{1,+1}$ ($\ket{1,0}\bra{1,-1}$), while the signal is intensified (attenuated) when $m_zN-n\bfM_z\cos\theta>0$. While the details of the terms containing the sum and difference of the spin-spin correlation functions $\chi_{nm}^{-+}$ and $\chi_{nm}^{+-}$ differ from system to system, it is a general conclusion that the intensity of the signal from any specific transition depends on the magnetic densities of the tip and substrate, and on their relative orientation.

We finally notice that the conductance $dI_0/dV$ vanishes for $nN+\bfm\cdot\bfM=0$, which corresponds to having a half-metallic tip and substrate with their magnetic moments in anti-parallel alignment. In this set-up also the conductance $dI_1/dV=0$, which implies that the measured signal is generated solely by $dI_2/dV$. Moreover, as can be seen in Eq. (\ref{eq-I2}), this conductance only depends on the transverse components, i.e. the sum and difference of $\chi_{nm}^{-+}$ and $\chi_{nm}^{+-}$, since the term containing $\chi_{nm}^z$ is proportional to $nN+\bfm\cdot\bfM\, (=0)$. Hence, despite the presence of possible thermal noise, such a set-up would benefit from a very low current noise since most of the noise would be related to the spin fluctuations that are to be measured. This can be seen by identifying the spin-dependent current operator with $\delta \hat{I}(t)=T_1\bfS(t)\cdot\bfs$, where $\bfs=\sum_{\bfp\bfk\sigma\sigma'}\cdagger{\bfp}\bfsigma\cs{\bfk\sigma'}$. The current-current correlation function is then given by \cite{balatsky2002} $\av{\delta\hat{I}(t)\delta\hat{I}(t')}=T_1^2\bfs\cdot\av{\bfS(t)\bfS(t')}\cdot\bfs$, where we average over the dynamics of the localized spins and over the ensemble of the tunneling electrons. Under the condition that $nN+\bfm\cdot\bfM=0$, the total dc current $I$ is  proportional to $T_1^2\int_{-\infty}^t\bfs\cdot\av{\bfS(t)\bfS(t')}\cdot\bfs\, d(t-t')$, and since the shot noise is approximately $\av{I_\text{shot}^2(\omega)}\sim I$,  the signal-to-noise ratio is about unity.

In the above examples we have, for the sake of argument, treated the local spins in the atomic limit and omitted mutual influences between the (electronic and magnetic) densities in the substrate and the adatoms. In reality, the adatom and the substrate affect one another and self-consistently creates a total effective e.g. magnetic moment locally around the adatoms. The basic formula given in Eq. (\ref{eq-I2}) remain unchanged, however, by a self-consistent treatment of the system.

The author thanks the Swedish Research Council (VR) and the Royal Swedish Academy of Sciences (KVA) for financial support. Special thanks to A. V. Balatsky, A. Bergman, S. Bl\"ugel, M. Bode, O. Eriksson, B. Gy\"orffy, A. Lichtenstein, L. Nordstr\"om, A. Taroni, and R. Wiesendanger for valuable discussions and critical comments, and to W. Wulfhekel for communicating results prior to publication.


\begin{thebibliography}{99}
\bibitem{jaklevic1966} R. C. Jaklevic and J. Lambe, Phys. Rev. Lett. {\bf 17}, 1139 (1966).
\bibitem{park2000} H. Park, J. Park, A. K. L. Lim, E. H. Andersson, A. P. Alivisatos, and P. McEuen, Nature, {\bf 407}, 57 (2000); 
R. H. M. Smit, Y. Noat, C. Untiedt, N. D. Lang, M. C. van Hemert, and J .M. van Ruitenbeek, \emph{ibid}, {\bf 419}, 906 (2002).
\bibitem{gorelik2001} L. Y. Gorelik, A. Isacsson, Y. M. Galperin, R. I. Shekter, and M. Jonson, Nature, {\bf 411}, 454 (2001); 
J. Fransson, J. -X. Zhu, and A. V. Balatsky, Phys. Rev. Lett. {\bf 101}, 067202 (2008).
\bibitem{gawronski2008} H. Gawronski, M. Melhorn, and K. Morgenstern, {\bf 319}, 930 (2008); 
J. Fransson and A. V. Balatsky, Phys. Rev. B, {\bf 75}, 195337 (2007).

\bibitem{gambardella2003} P. Gambardella, S. Rusponi, M. Veronese, S. S. Dhesi, C. Grazioli, A. Dallmeyer, I. Cabria, R. Zeller, P. H. Dederichs, K. Kern, C. Carbone, and H Brune, Science, {\bf 300}, 1130 (2003).
\bibitem{heinrich2004} A. J. Heinrich, J. A. Gupta, C. P. Lutz, and D. M. Eigler, Science, {\bf 306}, 466 (2004).
\bibitem{hirjibehedin2006} C. F. Hirjibehedin, C. P. Lutz, and A. J. Heinrich, Science, {\bf 312}, 1021 (2006).
\bibitem{hirjibehedin2007} C. F. Hirjibehedin, C. -Y. Lin, A. F. Otte, M. Ternes, C. P. Lutz, B. A. Jones, and A. J. Heinrich, Science, {\bf 317}, 1199 (2007).
\bibitem{balashov2008} T. Balashov, T. Schul, A. F. Tak\'acs, A. Ernst, S. Ostanin, J. Henk, I. Mertig, P. Bruno, T. Miyamachi, S. Suga, and W. Wulfhekel, arXiv:0903.3337v1.

\bibitem{balashov2006} T. Balashov, A. F. Tak\'acs, W. Wulfhekel, and J. Kirschner, Phys. Rev. Lett. {\bf 97}, 187201 (2006).
\bibitem{wahl2007} P. Wahl, P. Simon, D. Diekh\"oner. V. S. Stepanyuk, P. Bruno, M. A. Schneider, and K. Kern, Phys. Rev. Lett. {\bf 98}, 056601 (2007).
\bibitem{yayon2007} Y. Yayon, V. W. Brar, L. Senapati, S. C. Erwin, and M. F. Crommie, Phys. Rev. Lett. {\bf 99}, 067202 (2007).
\bibitem{meier2008} F. Meier, L. Zhou, J. Wiebe, and R. Wiesendanger, Science, {\bf 320}, 82 (2008).
\bibitem{chen2008} X. Chen, Y. -S. Fu, S. -H. Ji, T. Zhang, P. Cheng, X. -C. Ma, X. -L. Zou, W. -H. Duan, J. -F. Jia, and Q.-K. Xue, Phys. Rev. Lett. {\bf 101}, 197208 (2008).

\bibitem{fransson2008} J. Fransson, Nanotechnology, {\bf 19}, 285714 (2008); 
M. Persson, arXiv:0811.2511v1.
\bibitem{fransson2009} J. Fransson, O. Eriksson, and A. V. Balatsky, arXiv:0812:4956v2.
\bibitem{rossier2009} J. Fern\'andez-Rossier, arXiv:0901.4839v1.

\bibitem{balatsky2002} A. V. Balatsky, Y. Manassen, and R. Salem, Phys. Rev. B, {\bf 66}, 195416 (2002); 
J. -X. Zhu and A. V. Balatsky, Phys. Rev. B, {\bf 67}, 174505 (2003); 
Z. Nussinov, M. F. Crommie, and A. V. Balatsky, Phys. Rev. B, {\bf 68}, 085402 (2003).
\bibitem{bhattacharjee1992} A. K. Bhattacharjee, Phys. Rev. B, {\bf 46}, 5266--5273 (1992).
\bibitem{SSassumption} Eq. (\ref{eq-I2}) is written under the assumption that the time-dependent spin-spin correlation function e.g. $\chi^{-+}(t,t')$ can be replaced by $\chi^{-+}(t-t')$. Relaxing this assumption leads to a slightly more complicated formula which is not necessary for the present discussion.
\bibitem{tersoff1983} J. Tersoff and D. R. Hamann, Phys. Rev. Lett. {\bf 50}, 1998 (1983).
\bibitem{reittu1997} H. J. Reittu, J. Phys.: Condens. Matter, {\bf 9}, 10651 (1997).
\bibitem{wortmann2001} D. Wortmann, S. Heinze, Ph. Kurz, G. Bihlmayer and S. Bl\"ugel, Phys. Rev. Lett., {\bf 86}, 4132 (2001).
\end{thebibliography}
\end{document}